\documentclass[aps]{revtex4}  % for review 
\usepackage{graphicx}  % needed for figures
\usepackage{dcolumn}   % needed for some tables
\usepackage{bm}        % for math
\usepackage{amssymb}   % for math
\def\MET{{\mbox{$E\kern-0.57em\raise0.19ex\hbox{/}_{T}$}}}
\def\met{{\mbox{$E\kern-0.57em\raise0.19ex\hbox{/}_{T}$}}}
\def\DZero{D\O\ }
\def\Dzero{D\O\ }

\def\etal{{\it et.~al.}}

\begin{document}

\title{
Combination of CDF and \DZero Limits on a Gauge Mediated SUSY Model
Using Diphoton and Missing Transverse Energy Channel}

\author{V. Buescher}
\author{R. Culbertson}
\author{J. Conway}
\author{Y. Gershtein}
\author{J-F. Grivaz} 
\author{B. Heinemann} 
\author{D.H. Kim}
\author{M.S. Kim}
\author{S. Lammel}
\author{G. Landsberg}
\author{S.W. Lee}
\author{S. Mrenna}
\author{D. Toback}
\author{S.M. Wang}
\affiliation{\vskip0.1in for the CDF and \Dzero Collaborations }

\begin{abstract}
We combine the results of the CDF and \Dzero
searches for chargino and neutralino production
in Gauge-Mediated SUSY using the two-photon and
missing $E_T$ channel.  The data are $p \bar{p}$ collisions
produced at the Tevatron with $\sqrt{s}=1.96$~TeV,
with 202~pb$^{-1}$ collected at CDF and 263~pb$^{-1}$
collected at \Dzero.
The combined limit excludes a chargino mass
less than 209~GeV$/c^2$.  This result significantly extends the
individual experimental limits.
\end{abstract}

\maketitle

%%%%%%%%%%%%%%%%%%%%%%%%%%%%%%%%%%%%%%%%%%%%%%%%%%%%%%%%%%%%%%%%%%%%%%%
%%%%%%%%%%%%%%%%%%%%%%%%%%%%%%%%%%%%%%%%%%%%%%%%%%%%%%%%%%%%%%%%%%%%%%%
\section{Introduction} %%%%%%%%%%%%%%
Both CDF\cite{CDF} and \Dzero\cite{DZero} 
have reported on the search for an excess of events
containing two high-$p_T$ photons and significant
missing transverse energy.  The results have been
interpreted in the framework of a model 
of Gauge-Mediated Supersymmetry-Breaking (GMSB) with a
neutralino as the next-to-lightest supersymmetric particle.

The details of the model were chosen at the Snowmass Workshop\cite{snowmass}.
The complexity of a GMSB model is reduced to a one--parameter 
model, or model--line, that is qualitatively representative
of the phenomenology.  The model--line is defined as a function of a 
single mass parameter, $\Lambda$, which is varied to scan the masses of the particles 
in the model.  
At the Tevatron, this model allows for a significant production of 
the lightest chargino and second-lightest neutralino.  
These particles undergo a cascade decay to the lightest neutralino,
which itself decays to a photon and a gravitino.
The searches require two identified photons and large missing $E_T$.

Since the two experiments investigated the same model--line, 
the mechanics of the combination are straightforward.  The details
are explained in Section IV.  
As the cross sections and detection efficiencies vary along
the model--line,
the particle mass limits are 
set at the highest mass where the model can be excluded.
We will report the limits on chargino mass and $\Lambda$.
For clarity, we will discuss the details of the results for a chargino mass of 
approximately 200~GeV/$c^2$, which is near the combined limit,
and summarize information for other mass points.
The experiments chose different arbitrary mass points to measure efficiencies, 
so the CDF measurements have been interpolated to the \DZero masses.
In the following sections, we report the information that 
enters the combination: efficiency, luminosity,  
background expectations and data observations.  

%%%%%%%%%%%%%%%%%%%%%%%%%%%%%%%%%%%%%%%%%%%%%%%%%%%%%%%%%%%%%%%%%%%%%%%
%%%%%%%%%%%%%%%%%%%%%%%%%%%%%%%%%%%%%%%%%%%%%%%%%%%%%%%%%%%%%%%%%%%%%%%
\section{Efficiency and Luminosity} %%%%%%%%%%%%%%

%%%%%%%%%%%%%%%%%%%%%%%%%%%%%%%%
\subsection{CDF}

The CDF analysis requires that each event has a reconstructed interaction
vertex, the absolute value of the vertex $z$ is less than 
60~cm and the event passes the one of the diphoton triggers.
Both photons must be reconstructed in the central detector ($|\eta|<1$)
with $E_T>13$~GeV and pass fiducial cuts.
For isolation CDF requires $E_T$ in a $\eta-\phi$ cone of 0.4 to be 
less than  $0.1E_T$ if $E_T<20$~GeV, 
and $2+0.02E_T$ if $E_T>20$~GeV.
Photon identification also includes requirements on 
tracking isolation, the ratio of hadronic to electromagnetic energy,
and shower shape.
The topological cuts require that the \met~ does not point 
along or opposite to 
a jet and that there is no evidence of a cosmic ray or 
beam--related accidental energy deposition.  The \met~ cut is 45~GeV.
The total acceptance times efficiency for signal events 
with a chargino mass of 200~GeV$/c^2$ is 7.3\%.

The systematic uncertainty on the efficiency of the photon 
identification cuts is 13\%, determined from the variation of the result 
using different techniques.  
Varying the parton distribution functions (PDF) causes a 5\% 
change in the efficiency.  Varying the initial-- and final--state 
radiation implies a 10\% systematic uncertainty 
and varying the hard scale $Q^2$ gives a 3\% systematic uncertainty 
on the acceptance times efficiency.

The analysis includes $202\pm 12$~pb$^{-1}$, where the uncertainty 
is systematic, coming from the inelastic cross section and the total 
detector acceptance for inelastic events.

%%%%%%%%%%%%%%%%%%%%%%%%%%%%%%%%
\subsection{\Dzero}

The \DZero analysis requires that both photons are reconstructed 
in the central calorimeter ($|\eta|<1.1$) 
and have $E_T > 20~$ GeV.  
The events are triggered by a combination of single--cluster and 
two--cluster electromagnetic triggers. 
The events are required
to have a reconstructed vertex and not to exhibit patterns of
correlated calorimeter noise.
Photon clusters
are selected from all calorimeter clusters by requiring that: (i) at 
least 90\% of the energy is deposited in the EM section of the calorimeter,
(ii) the calorimeter isolation variable 
$I=[E_{tot}(0.4)-E_{EM}(0.2)]/E_{EM}(0.2)$ is less than 0.15, where
$E_{tot}(0.4)$ is the total energy in a cone of radius 0.4 in 
$\eta - \phi$ space and $E_{EM}(0.2)$ is the EM energy in a cone of 
radius 0.2, (iii) the transverse and longitudinal shower profiles
are consistent with those expected for an EM shower, and (iv) the
scalar sum of the $P_T$ of all tracks in annulus of 
$0.05 < {\cal R} < 0.4$ around the cluster is less than 2 GeV$/c$.
Finally, to reject electrons, photon candidates are required to have
no central track well-matched to the cluster.

The efficiency for photon reconstruction and identification was
obtained by reconstructing Monte Carlo events passed through a detailed GEANT
simulation of the \DZero detector. The simulation was verified by
comparing to $Z \rightarrow e^+e^-$ data. 
Special attention has been paid to efficiency dependence on the
number of jets in the event and on the distance between the electron
and the closest jet. The estimated systematic uncertainty on total
efficiency from photon acceptance and identification cuts is 8\%.

The \MET~ in the event is required to be larger than 40~GeV.  The event 
must also pass topological cuts on the direction of
\MET, namely that it is not opposite to the leading jet (if present)
or along any of the photons.

The total efficiency for a chargino mass of 196~GeV$/c^2$ 
is 14.9\%. In addition to
the acceptance and photon identification uncertainty, there is 
 an additional systematic uncertainty coming from
the choice of PDF (5\%) and Monte Carlo statistics (4\%).

The \Dzero analysis includes $263\pm 17$~pb$^{-1}$, where the uncertainty 
is systematic, coming from the inelastic cross section and 
the detector acceptance for inelastic events.

%%%%%%%%%%%%%%%%%%%%%%%%%%%%%%%%
\subsection{Correlations}

The systematic uncertainties on the efficiencies 
of photon identification variables are assumed to be
uncorrelated since the two analyses are based on different detectors, cuts
and methods.
Since the 5\% PDF systematic originates from the same source for 
both experiments, and reflects effects such as changing 
particle $P_T$, which would affect both experiments similarly, 
the 5\% PDF systematics are assumed 100\% correlated.  
Since both experiments base their luminosity estimates on the 
same inelastic cross section measurement, and both base their acceptance
on the same Monte Carlo generator, the luminosity systematics are 
assumed to be 100\% correlated.
The uncorrelated CDF uncertainty on acceptance times efficiency is 17\%.
The corresponding uncorrelated \DZero uncertainty is 9\%.

%%%%%%%%%%%%%%%%%%%%%%%%%%%%%%%%%%%%%%%%%%%%%%%%%%%%%%%%%%%%%%%%%%%%%%%
%%%%%%%%%%%%%%%%%%%%%%%%%%%%%%%%%%%%%%%%%%%%%%%%%%%%%%%%%%%%%%%%%%%%%%%
\section{Background and Data Observations} %%%%%%%%%%%%%%

%%%%%%%%%%%%%%%%%%%%%%%%%%%%%%%%
\subsection{CDF}

CDF considers the following sources of backgrounds.
The background from photons and jets faking photons with fake \met~ 
is estimated to be $0.01\pm 0.01{\rm (stat)} \pm 0.01{\rm (syst)}$ events 
and is small enough to ignore in the combination.
The background from events with a true electron and a real or fake photon, 
where the electron then fakes a photon is 
$0.14 \pm 0.06{\rm (stat)} \pm 0.05{\rm (syst)}$ events.
The systematic uncertainty is from the uncertainty in 
the purity of the electron in the $e\gamma$ sample.
The background from non--collision sources is 
$0.12 \pm 0.03{\rm (stat)} \pm 0.09{\rm (syst)}$ events.
The total background is $0.27\pm0.07{\rm (stat)}\pm 0.10{\rm (syst)}$ events.
The total statistical and systematic uncertainty is then 12\%.

CDF observed no events passing all cuts.

%%%%%%%%%%%%%%%%%%%%%%%%%%%%%%%%
\subsection{\Dzero}

\DZero considers two types of backgrounds. Background from QCD events with
either real or fake photons and mis-measured \MET~ is estimated to be
$2.8 \pm 0.5$ events, with uncertainty dominated by statistics in the
sample used for the estimate. The background from events with an
electron mis-identified as a photon is $0.9 \pm 0.2$ events,
with an uncertainty dominated by statistics.
The total background is $3.7 \pm 0.6$ events.

\DZero observed 2 events passing all cuts.

%%%%%%%%%%%%%%%%%%%%%%%%%%%%%%%%
\subsection{Correlations}

Since only \DZero has a significant QCD background, its uncertainty 
is uncorrelated with CDF.
The systematic uncertainty on the $e\gamma$ background 
is considered to be uncorrelated since \DZero is dominated by statistics.
The background from non--collision sources is 
negligible in \Dzero and the corresponding uncertainty 
is therefore not correlated.

%%%%%%%%%%%%%%%%%%%%%%%%%%%%%%%%%%%%%%%%%%%%%%%%%%%%%%%%%%%%%%%%%%%%%%%
%%%%%%%%%%%%%%%%%%%%%%%%%%%%%%%%%%%%%%%%%%%%%%%%%%%%%%%%%%%%%%%%%%%%%%%
\section{Combination} %%%%%%%%%%%%%%

The combination proceeds using the data in Table \ref{tab:combo} and the 
prescription from \cite{bib:john}.
The method forms a Bayesian likelihood from the product of likelihoods 
of the individual experiments, with flat priors.  
Each correlated and uncorrelated 
systematic uncertainty is represented by an appropriate 
Gaussian function.  
Finally we integrate over all parameters except the cross section,
and integrate the cross section to the 95\% confidence level point.

The table also includes the expected limit for the experiments and 
the combination.  The expected limit is found by computing the limit for each 
possible outcome, given the expected background, and taking the average, 
weighted by the probability of that outcome.

\begin{table}[h]\begin{center}
\begin{tabular}{|c|c|c|c|c|c|} \hline\hline
\multicolumn{6}{|c|}{CDF}\\ \hline
$\chi^\pm_1$ Mass (GeV$/c^2$) & 154 & 168  & 182 & 196 & 209 \\ \hline
$\epsilon$ (\%) & 6.4 & 6.8 & 7.2 & 7.4 & 7.6  \\ \hline
$\sigma(\epsilon)/\epsilon$ (\%) uncorrelated  & \multicolumn{5}{c|}{17} \\ \hline
$\sigma(\epsilon)/\epsilon$ (\%) correlated, from PDF & \multicolumn{5}{c|}{5} \\ \hline
${\cal L}$ (pb)                   & \multicolumn{5}{c|}{202}  \\ \hline
$\sigma({\cal L})/{\cal L}$ (\%) correlated   & \multicolumn{5}{c|}{6}  \\ \hline
$b$ (events)               & \multicolumn{5}{c|}{0.27}  \\ \hline
$\sigma(b)$ (events) uncorrelated & \multicolumn{5}{c|}{0.12}  \\ \hline
observed events                   & \multicolumn{5}{c|}{0}  \\ \hline
CDF cross section$\times$BR$^2$  limit (pb)  & ~0.254~ & ~0.239~ & ~0.225~ & ~0.219` & ~0.213~ \\ \hline
CDF cross section$\times$BR$^2$  expected limit (pb)  & 0.294 & 0.277 & 0.261 & 0.254 & 0.247 \\ \hline\hline
\multicolumn{6}{|c|}{\Dzero}\\ \hline
$\chi^\pm_1$ Mass (GeV$/c^2$) & 154 & 168  & 182 & 196 & 209 \\ \hline
$\epsilon$ (\%) & 11.1 & 12.4 & 13.7 & 14.9 & 15.4 \\ \hline
$\sigma(\epsilon)/\epsilon$ (\%) uncorrelated   & \multicolumn{5}{c|}{9} \\ \hline
$\sigma(\epsilon)/\epsilon$ (\%) correlated PDF & \multicolumn{5}{c|}{5} \\ \hline
${\cal L}$ (pb)                   & \multicolumn{5}{c|}{263}  \\ \hline
$\sigma({\cal L})/{\cal L}$ (\%) correlated   & \multicolumn{5}{c|}{6.5} \\ \hline
$b$ (events)               & \multicolumn{5}{c|}{3.7} \\ \hline
$\sigma(b)$ (events) uncorrelated & \multicolumn{5}{c|}{0.6} \\ \hline
observed events                   & \multicolumn{5}{c|}{2}  \\ \hline
\DZero cross section$\times$BR$^2$ limit (pb)   & 0.153 & 0.137 & 0.124 & 0.114 & 0.110 \\ \hline
\DZero cross section$\times$BR$^2$ expected limit (pb)   & 0.214 & 0.192 & 0.174 & 0.160 & 0.154 \\ \hline\hline
\multicolumn{6}{|c|}{Combined CDF and \Dzero}\\ \hline
$\chi^\pm_1$ Mass (GeV$/c^2$) & 154 & 168  & 182 & 196 & 209 \\ \hline
cross section$\times$BR$^2$ limit (pb)      & 0.099 & 0.090 & 0.083 & 0.077 & 0.075 \\ \hline
cross section$\times$BR$^2$ expected limit (pb)      & 0.141 & 0.129 & 0.118 & 0.111 & 0.108 \\ \hline
LO model cross section$\times$ BR$^2$ (pb) & 0.317 & 0.202 & 0.143 & 0.094 & 0.068 \\ \hline
NLO model cross section$\times$ BR$^2$ (pb)           & 0.349 & 0.224 & 0.159 & 0.106 & 0.077 \\ \hline\hline
\end{tabular}
\end{center}
\caption{The numbers used in the combined limits. 
The branching ratio for the lightest neutralino to decay to a photon and 
gravitino is 0.95. The symbols $\epsilon$, ${\cal L}$,  $b$, and $\sigma(x)$ 
represent acceptence times efficiency, 
integrated luminosity, predicted background event counts, 
and the uncertainty on $x$, respectively.}

\label{tab:combo}
\end{table}

%%%%%%%%%%%%%%%%%%%%%%%%%%%%%%%%%%%%%%%%%%%%%%%%%%%%%%%%%%%%%%%%%%%%%%%
%%%%%%%%%%%%%%%%%%%%%%%%%%%%%%%%%%%%%%%%%%%%%%%%%%%%%%%%%%%%%%%%%%%%%%%
\section{Limit} %%%%%%%%%%%%%%

    The combined analyses  set a limit on the  a total
production cross section for supersymmetric particles
with the decay of the lightest neutralino into a photon
and a gravitino.  
The cross section limit is interpreted as a chargino mass
for a point along the model--line described above.

\begin{figure}[ht]
\begin{centering}
\includegraphics[width=12.0cm]{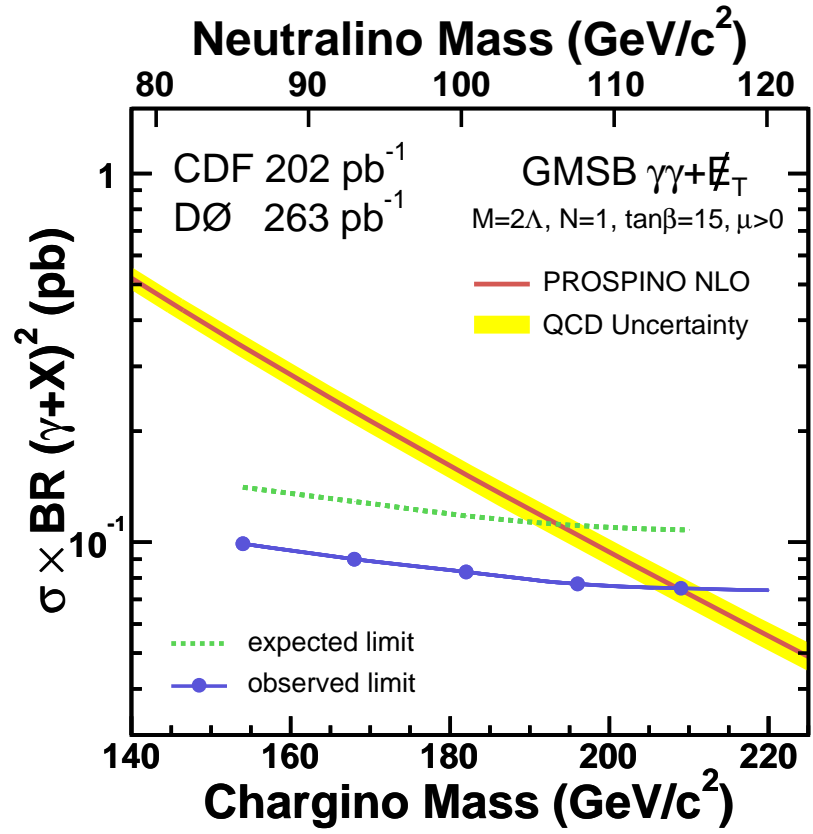}
\caption{
\label{fig:xsec}
The next-to-leading-order cross section and 
combined experimental limits as a function of chargino 
and neutralino mass.
}
\end{centering}
\end{figure}

The branching ratio for the lightest neutralino to decay to a 
photon and gravitino is computed by 
ISAJET V7.51\cite{bib:ISAJET} and is included in
the limit setting process. This has a value of approximately 0.95 at a
chargino mass of 200~GeV$/c^2$.  
The LO and NLO production cross sections were computed with 
Prospino 2.0\cite{bib:prospino} using the same GMSB parameters.
The cross sections 
and the cross section limits as a function of chargino mass 
are displayed in Fig. \ref{fig:xsec}. 
The final mass limit for the lightest chargino is
209~GeV$/c^2$ which translates to a mass limit of 114~GeV$/c^2$ 
on the lightest neutralino and a limit of 84.6~TeV on $\Lambda$.
This result improves 
significantly on the mass limits of the individual experiments, which
exclude charginos with a mass below 195~GeV$/c^2$ (\Dzero) 
and 167~GeV$/c^2$ (CDF), both derived using slightly different predictions for
cross section times branching ratio.

%%%%%%%%%%%%%%%%%%%%%%%%%%%%%%%%%%%%%%%%%%%%%%%%%%%%%%%%%%%%%%%%%%%%%%%
%%%%%%%%%%%%%%%%%%%%%%%%%%%%%%%%%%%%%%%%%%%%%%%%%%%%%%%%%%%%%%%%%%%%%%%

\clearpage

\end{document}